%% file: paper.tex
\documentclass[runningheads]{llncs}

\input{packages}
\input{commands}
\begin{document}

\title{Compensating the Packet Delay Variation for 6G Integrated with IEEE Time-Sensitive Networking}
\titlerunning{Packet Delay Variation Compensation in 6G Integrated with IEEE TSN}

\author{
  Marilet De Andrade \inst{1}\orcidID{0000-0002-0933-0167} \and
  Joachim Sachs\inst{2}\orcidID{0000-0001-8398-7741} \and
  Lucas Haug\inst{3}\orcidID{0009-0003-7800-0836} \and
  Simon Egger\inst{3}\orcidID{0009-0007-9583-6044} \and
  Frank Dürr\inst{3}\orcidID{0000-0002-3470-7712} \and 
  Balázs Varga\inst{4}\orcidID{0000-0002-8574-330X} \and
  Janos Farkas\inst{4}\orcidID{0000-0001-5277-8402} \and
  György Miklós \inst{4}
} 

\authorrunning{M. De Andrade et al.}

\institute{
    Ericsson, 16440 Stockholm, Sweden 
    \email{marilet.de.andrade.jardim@ericsson.com}
\and 
    Ericsson, 52134 Aachen, Germany,
    \email{joachim.sachs@ericsson.com}
\and
    University of Stuttgart, Germany,
    \email{\{lucas.haug,simon.egger,frank.duerr\}@ipvs.uni-stuttgart.de}
\and 
    Ericsson, 1117 Budapest, Hungary,
    \email{\{balazs.a.varga,janos.farkas, gyorgy.miklos\}@ericsson.com}
}

\maketitle              
\begin{abstract}
  6G is deemed as a key technology to support emerging applications with stringent requirements for highly dependable and time-critical communication.
  In this paper, we investigate 6G networks integrated with TSN and how to compensate for wireless stochastic behavior which involves a large intrinsic packet delay variation.
  We evaluate a 6G solution to reduce packet delay variation that is based on de-jittering.
  For this, we propose to use virtual timeslots for providing the required time-awareness.
  We discuss the benefits of the proposed solution while evaluating the impact of the timeslot size on the number of schedulable TSN streams.

\end{abstract}

\input{content/01_introduction}

\input{content/03_related_work}
\input{content/02_background}
\input{content/04_design}
\input{content/05_evaluation}
\input{content/06_discussion}
\input{content/07_conclusion}

\begin{credits}

\subsubsection{\ackname} This work was supported by the European Union’s Horizon Europe project DETERMINISTIC6G under grant agreement No. 101096504.

\subsubsection{\discintname}
The authors have no competing interests to declare that are relevant to the content of this article.
\end{credits}

\printbibliography

\end{document}

%% file: packages.tex
\usepackage[T1]{fontenc}
\usepackage{color}
\usepackage{soul}

\usepackage{amsmath,amssymb,amsfonts}
\usepackage{graphicx}
\usepackage{textcomp}
\usepackage{xcolor}
\usepackage{booktabs}
\usepackage{dblfloatfix}
\usepackage{cancel}
\usepackage{url}

\usepackage{algorithm}
\usepackage{algpseudocode}

\usepackage[inline]{enumitem}
\usepackage{subcaption}
\usepackage{siunitx}
\usepackage{biblatex}

\usepackage{tikz}
\usepackage{pgfplotstable}
\pgfplotsset{compat=1.16}
\usepgfplotslibrary{fillbetween}

\usetikzlibrary{patterns, patterns.meta} 
\usetikzlibrary{calc}
\usetikzlibrary{positioning}
\usetikzlibrary{arrows.meta}
\usetikzlibrary{math} 
\usetikzlibrary{backgrounds,scopes}
\usetikzlibrary{pgfplots.statistics, pgfplots.colorbrewer} 
\usetikzlibrary{trees} 
\usetikzlibrary {datavisualization.formats.functions}
\usetikzlibrary{fit}
\usetikzlibrary{shapes.geometric}
\usetikzlibrary{shadows}
\usetikzlibrary{shapes}
\usetikzlibrary{external}

\usepackage{algpseudocode}

\usepackage[newfloat,frozencache,cachedir=minted-cache]{minted}
\setminted{
	fontsize={\scriptsize},
	linenos=false,
	stripall=true,
	breaklines=true,
	autogobble=true,
	xleftmargin=0.5em,
	texcomments,
	escapeinside=~~,
	tabsize=2,
	frame=single
}

\usepackage{comment}


%% file: commands.tex
\definecolor{todoyellow}{RGB}{255,255,120}
\sethlcolor{todoyellow}

\newcommand{\thinskip}[0]{\hspace*{0.16667em}\relax}
\newcommand{\jrdash}[2]{\unskip #1\thinskip #2\thinskip\ignorespaces}

\newcommand{\ldash}[0]{\jrdash{\empty}{\hbox{---}\nobreak}}
\newcommand{\rdash}[0]{\jrdash{\nobreak}{---}}

%% file: content/01_introduction.tex
\section{Introduction}

Future 6G networks are expected to play an important role in offering dependable and time-critical connectivity service. Such a service will be crucial for many emerging use cases such as adaptive manufacturing and mobile automation, extended reality, and occupational exoskeletons, as envisioned in \cite{det6g_usecases}. These use cases require stringent guarantees on Packet Delay (PD) and Packet Delay Variation (PDV). Enabling these capabilities in 6G will be crucial to fulfill the requirements of these use cases. Time Sensitive Networking (TSN) is a layer 2 technology that guarantees end-to-end bounded latency and very low packet losses in Ethernet networks. 3GPP has standardized the integration of 5G and TSN in order to support wireless delay-critical use cases. TSN support is also is expected to be supported in the next generation of mobile communications, 6G with additional enhancements to cover for specific needs. An important issue that requires a solution for 6G is the significant PDV introduced by the wireless TSN bridge (the 6G system acting as a TSN bridge). 5G/6G exhibits a larger packet delay variation (PDV) compared to a regular wired TSN bridge. The measurements shown in \cite{downlink_example_histogram} show that the 5G system presents a PDV in the order of milliseconds, while a wired bridge shows a PDV in the order of 100s of nanoseconds. Such a large PDV poses a challenge for a TSN scheduler in calculating an end-to-end schedule for the TSN streams, which in turn makes on-time data delivery difficult.

Also for many delay-critical applications, a large jitter (or PDV) is not acceptable. For example, in manufacturing and industrial automation, the application sensitivity towards jitter has been noted in Table 1 of the 5G-ACIA white paper on 5G integration with TSN~\cite{5g_acia}, where multiple traffic types show little or no tolerance to PDV. Therefore, a solution is needed to remove PDV or compensate for it. If a resulting PDV is in the range of microseconds rather than milliseconds, existing TSN scheduling algorithms can be applied, without causing undesired packet disordering as described later. Since the main contributor to PDV in the end-to-end path is the wireless link, we focus on PDV compensation within the 6G system to reduce the stochastic uncertainty in packet delivery. We denote this method as packet delay correction (PDC). An assessment of the newly proposed PDC mechanism is provided.

The structure of the paper starts with a background discussion in section 2, and then we analyze the related work in section 3. Section 4 centers the attention on packet delay correction mechanisms and how they are applied. Simulation-based evaluations of the performance of a TSN network with a wireless TSN bridge are presented in section 5. Section 6 concludes the paper.

%% file: content/03_related_work.tex
\section{Related Work}

The proposed mechanism is intended to compensate for the 5G/6G PDV, which is using a similar concept as in Logical Execution Time (LCE) where variable processing time is eliminated by setting the end of the task near the end of the task interval or task deadline. LCE was originally proposed in~\cite{1173196} and the works related to LCE for determinism have been summarized in~\cite{wang2024optimizinglogicalexecutiontime}. 

Existing 5G-TSN literature focuses on TSN mitigation strategies to compensate the 5G packet delay variations.
We argue that existing work falls into either of the categories described in the following sub-sections.

\subsection{TSN Scheduling}
There exists a vast literature on synthesizing suitable IEEE 802.1Qbv Gate Control Lists (GCL, i.e. the resulting TSN schedule for a bridge) to provide bounded latency and jitter guarantees in wired networks;
a recent survey can be found in~\cite{10151872}.
They explore different TSN scheduler designs, ranging from exact solvers that optimize a certain objective~\cite{8894249,Craciunas2016RTNS} to schedulers that prioritize speed over optimality by employing heuristics or machine learning~\cite{nwps,VLK2022105512}.
Moreover, there exist numerous extensions to account for joint scheduling and routing~\cite{10.1145/3139258.3139289}, multicast paths~\cite{9203662}, or dynamic schedule adaptation at runtime~\cite{8607244} to name just a few.
However, all of them have in common that they are designed for wired networks with deterministic transmission delays; they do not account for the significant 5G packet delay variations experienced in a 5G-TSN setting. 

More recent work aims to remedy this shortcoming by introducing designated TSN scheduling constraint systems~\cite{9212049,9940254,Egger2025}.
The authors of~\cite{9212049,9940254} consider a joint optimization with 5G-internal resource allocation to synthesize TSN schedules. 
However, they still assume constant 5G delay characteristics that result in similar delay assumptions as the deterministic models of wired TSN.
In contrast, \cite{Egger2025} introduces a wireless-friendly TSN scheduler that utilizes 5G packet delay budgets to explicitly capture 5G packet delay variations up to the required reliability percentile.
While the authors formally analyze that their QoS guarantees are upheld end-to-end, this paper aims to explore a different approach:
we design PDC as a complementary mechanism deployed in the 5G/6G system to enable conventional TSN schedulers from wired TSN without having to make them wireless-aware by modifying their underlying scheduling constraints. 

\subsection{Jitter Control Mechanisms}
We continue with the discussion on existing work that follows a more closely related design to compensate 5G packet delay variations:
Most prominently, 3GPP defines a hold-and-forward buffering mechanism~\cite{3gpp.23.501} at the egress ports of the 5G system.
However, this mechanism is only meant to provide ``externally observable behavior identical to scheduled traffic with eight queues''~\cite{3gpp.23.501} in adherence to IEEE 802.1Qbv and thereby crucially differs from the per-stream de-jittering capability that this paper introduces with PDC.
In particular, our evaluations show that this is a crucial difference to provide end-to-end QoS guarantees with conventional TSN schedulers in the 5G/6G-TSN network.

We also note that the concept of jitter control is not new but has a strong foundation in real-time communication literature.
For instance, a non-work-conserving queuing discipline capable of jitter control was already introduced by~\cite{253355} for generic connection-oriented networks.
It operates by maintaining a state (at each hop) on the eligibility time of packets and by ensuring a minimum delay between two subsequent packets of the same connection.
More recent work~\cite{9829777,Fontalvo2024} proposes similar TSN mechanisms based on the IEEE 802.1Qcr Asynchronous Traffic Shaper~\cite{802.1Qcr}. Also there has been numerous research works to control jitter by using TSN traffic shaper based on Cyclic Queueing and Forwarding (CQF), which related work was summarized and further investigated in~\cite{11122365}.
Usually the CQF is applied at a per-traffic-class level.
However, the problem in 5G/6G networks would require a CQF per TSN stream due to the uncertain queueing ordering problem (explained in Section 3), which may signify a large implementation complexity. 
In contrast, this work introduces 5G-specific PDC architectures to eliminate any 5G packet delay variation at its source to support the IEEE 802.1Qbv or Time-Aware Shaper (TAS); and this approach is equally relevant for future 6G networks. 

We consider the focus on IEEE 802.1Qbv or TAS as relevant since it is important or even mandatory for some applications, e.g., in manufacturing and industrial automation (see Table 1 in~\cite{5g_acia}).
We thereby aim to stay complementary to potentially deployed TSN mechanisms and, in particular, to enable the use of conventional TSN schedulers for TAS without requiring them to become ``wireless-aware''.

%% file: content/02_background.tex
\section{Background and Problem Statement}
5G has been standardized to support the integration with TSN networks. 6G is expected to adopt the support for TSN from 5G. This section provides background information on the most important 5G and TSN mechanisms.
Thereafter, we illustrate the problem of using conventional TSN scheduling and configuration mechanisms, as developed for wired TSN networks, when TSN is integrated with 5G and is exposed to significant packet delay variations.

\subsection{Introduction to 5G-TSN Integration}
IEEE 802.1 Time-Sensitive Networking (TSN) is a layer 2 standard technology designed to deliver deterministic services that guarantee bounded low latency and very low packet losses. 
TSN provides a toolset that consists of several features, including TSN scheduling~\cite{802.1Qbv}, per-stream filtering and policing~\cite{802.1Qci}, time synchronization~\cite{802.1AS}, and frame replication and elimination~\cite{802.1CB}.
For the scope of this work, we investigate the behavior of IEEE 802.1Qbv traffic scheduling (in the remainder of the paper simply referred to as TSN scheduling) in a setting where 5G or 6G is integrated into a TSN network to provide wireless TSN connectivity.
Each TSN bridge is equipped with up to eight FIFO queues per egress port that are controlled by a so-called Gate Control List~(GCL).
The GCL specifies the exact opening and closing times of the gate for each egress queue, which repeats in cycles (a.k.a. hypercycles). 
Thereby, the first frame in the egress queue is eligible for transmission if the queue's assigned gate is open long enough to complete the frame's transmission.\footnote{
For completeness, we note that for multiple queues, the transmission selection algorithm of IEEE 802.1Qbv selects the next eligible frame from all the queues with an open gate following the order of the priorities assigned to the queues.}
The central challenge of providing end-to-end QoS guarantees for each TSN stream is thus to compute suitable GCLs for all TSN bridges that accounts for all time-critical TSN streams and their transmission delays along their end-to-end paths.

As synthesizing such GCLs is a computationally complex task~\cite{10151872}, a typical TSN network controller is centralized and has a global view on the network and the streams.
Network control can be split into two components~\cite{802.1Qcc}:
First, the Central User Configuration (CUC) controller communicates with end hosts and defines the TSN streams and their requirements.
Second, the stream requirements are given to the Centralized Network Configuration (CNC) controller which then gathers information from the network nodes, to perform route and scheduling calculations, and to configure the ports of the bridge components in the network. 

\begin{figure}
    \centering
    \includegraphics[width=\linewidth]{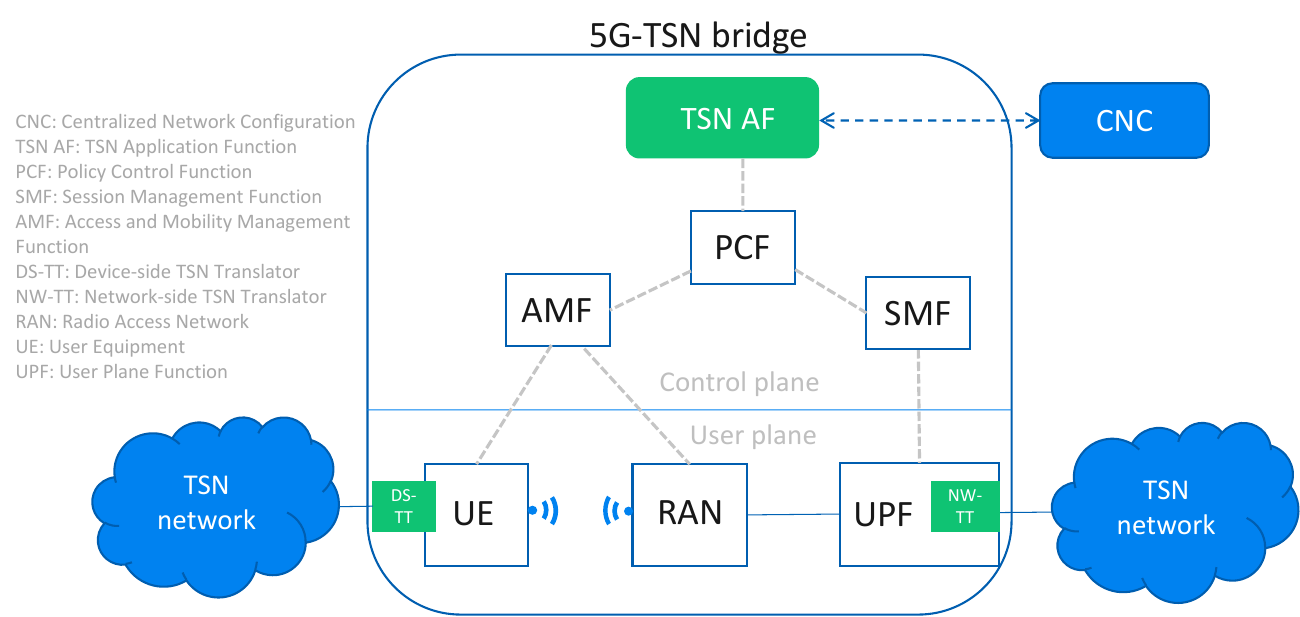}
    \caption{The 5G system modeled as a logical TSN bridge, turning into a 5G-TSN bridge, as standardized by 3GPP TS 23.501~\cite{3gpp.23.501}.}
    \label{fig:logicalbridge}
\end{figure}

More recent research and standardization efforts also aim to enhance support for TSN in 5G by modeling the 5G system as a logical TSN bridge.
3GPP has defined two data plane entities to support the integration with TSN~\cite{3gpp.23.501}: the device-side TSN translator (DS-TT) and the network-side TSN translator (NW-TT).
The DS-TT and the NW-TT support common TSN functionality (e.g., time synchronization, TSN scheduling, and PSFP) that allow the CNC to configure the 5G-TSN bridge like any other wired TSN bridge.
Furthermore, control-plane support for TSN has been added by introducing a TSN Application Function (TSN-AF) as the control-plane entity that maintains the interface between 5G and the CNC, as shown in Fig.~\ref{fig:logicalbridge}.
The TSN AF collects information from the 5G system and the DS-TTs and the NW-TT such as port capabilities, and minimum and maximum bridge delay, and then reports collected information to the CNC. Having the capabilities of every bridge and the requirements of every TSN stream, the CNC performs calculations and provides the configuration to every bridge and port.
Thereby, the TSN AF translates the TSN parameters and requirements \ldash as configured by the CNC \rdash into 5G parameters and triggers the configuration of the 5G connectivity accordingly. 

\subsection{Compensating Packet Delay Variation}
The PDV of the 5G system is significantly larger compared to delays encountered in wired TSN nodes. This is partly due to that an entire mobile network is represented towards the TSN domain as a single (logical) TSN node. In addition, several characteristics of wireless transmission cause delay variations, such as the radio frame structure for radio resource allocations and time-division duplexing, or fast retransmissions that efficiently provide highly reliable wireless transmission despite large signal variations caused by radio propagation in a dynamically changing radio environment. Even if 5G has specified a set of features for ultra-reliable and low latency communication (URLLC), this URLLC toolbox allows to configure the 5G system such that data can be reliably transmitted within a specified latency bound. However, URLLC only enables to provide an upper bound to the packet delay, it does not prohibit the large packet delay variations within this bound. Similar issues are expected with the next generation, 6G.

It is desirable to enable the 5G or 6G system to achieve PDV characteristics with low variation (e.g. down to tens of microseconds) so that they become closer to the PDV encountered in wired TSN bridges. This would allow to maintain the same TSN toolset used for planning and configuration of wired TSN networks, and apply it equally for TSN networks that are integrated with 5G/6G. 

\begin{figure}
    \centering
    \includegraphics[width=.6\linewidth]{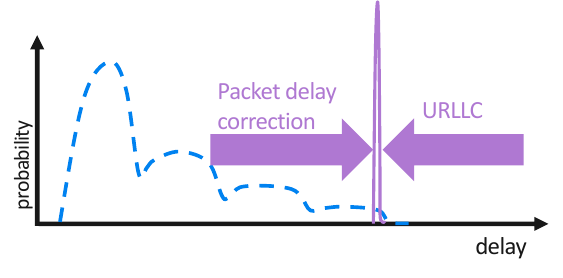}
    \caption{Bounded delay through PDC and URLLC.}
    \label{fig:pdc_urllc}
\end{figure}

A suitable approach to achieve low PDV in a 5G/6G network is to force packets that have been transmitted through the wireless network, to be passed on to the next TSN node right on time, i.e., not earlier and not later. We propose a packet delay correction (PDC) mechanism that ensures that the lower delay bound is just below the upper bound (see Fig. \ref{fig:pdc_urllc}). The resulting PDV would become very small, providing a high certainty to the CNC controller in planning the transmission times of TSN frames.

The possibility to  compensate delay variation in 5G was discussed in standardization in the 3GPP System Architecture working group SA2. However, the proposed hold-and-forward buffering mechanism (HF) was only defined to mimic the TSN scheduling gates at the ports (see clause 5.27.4 in \cite{3gpp.23.501}). That is, the frames are held when the corresponding gate is closed, and frames are forwarded when the gate is open. 3GPP decided to leave such a mechanism for implementation. What is missing in the HF mechanism is a de-jittering function that holds frames until reaching a pre-defined target delay, which ideally would be the same for all frames and leading to a zero jitter or PDV. To this end, a HF mechanism could be applied at the DS-TTs and NW-TT to pace out the frames to the next neighboring TSN bridge without any delay variation. The 5G system would then bound the PDV to stay within a controlled range.  

For a future 6G system, we consider it desirable to specify a common HF mechanism to provide compensation of the PDV such that time-critical communication with TSN is enabled.

\subsection{The Problem of Uncertain Queuing Orderings} \label{sec:uncertain_queueing}

Especially for TSN scheduling under IEEE 802.1Qbv, any deviation from the intended schedule can result in uncontrolled queuing backlogs and potentially nullify any end-to-end QoS guarantees.
While this problem is even encountered for small delay variations in wired TSN (e.g., due to microsecond time synchronization errors or sporadic frame loss~\cite{Craciunas2016RTNS}), these effects are amplified in 5G-TSN since 5G packet delay variations are in the range of multiple milliseconds~\cite{downlink_example_histogram}, a difference by three orders of magnitude.
Recall that the number of egress queues of TSN bridges is limited to 8, meaning that the TSN schedule has to account for packets that are arriving from different ingress ports and are mixed in the same egress queue.
Consequently, any uncertain reordering within the egress queues can result in packets missing their intended transmission slots and potentially having to wait until the next hypercycle for another transmission opportunity. 
These reordering effects may even propagate throughout the entire network and can lead to exacerbated end-to-end delays of time-critical streams.

\begin{figure}
    \centering
    \begin{subfigure}{\textwidth}
        \centering
        \includegraphics[width=\textwidth]{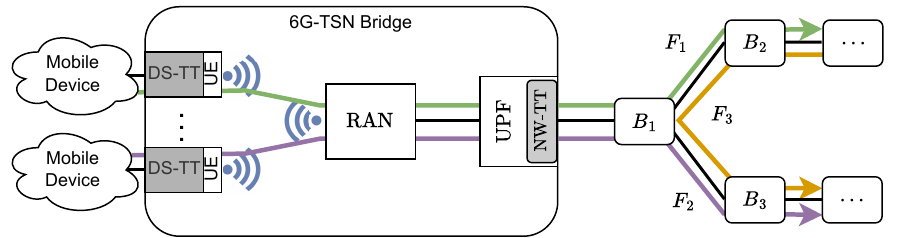}
        \caption{Network topology and stream routes}
        \label{fig:uncertain_queueing_problem:topo}
    \end{subfigure}
    \begin{subfigure}{0.48\textwidth}
        \centering
        \includegraphics[width=\textwidth]{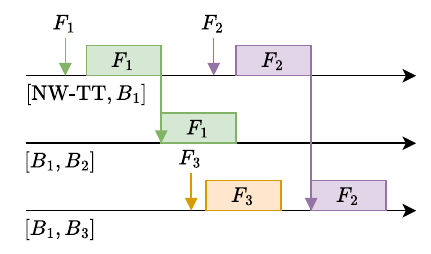}
        \caption{Intended Transmission Ordering}
        \label{fig:uncertain_queueing_problem:indended}
    \end{subfigure}
    \hfill 
    \begin{subfigure}{0.48\textwidth}
        \centering
        \includegraphics[width=\textwidth]{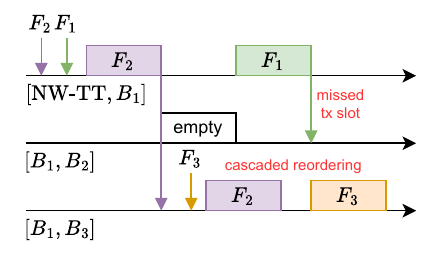}
        \caption{Faulty Transmission Ordering}
        \label{fig:uncertain_queueing_problem:faulty}
    \end{subfigure}
    \caption{Uncertain queuing orderings for streams $F_1$ and $F_2$. 
        Without accounting for the significant 6G packet delay variations, $F_2$ can arrive earlier than expected by the IEEE 802.1Qbv schedule.
    }
    \label{fig:uncertain_queueing_problem}
\end{figure}

Fig.~\ref{fig:uncertain_queueing_problem} provides a minimal example to illustrate the problem of uncertain queuing orderings.
The network topology and stream routes are shown in Fig.~\ref{fig:uncertain_queueing_problem:topo}, where we assume a single egress queue for the NW-TT and $B_1$ (for each of the two egress ports).
There are two wireless uplink streams ($F_1$ and $F_2$) and one internal streams in the wired TSN backbone ($F_3$).
The intended transmission slots for $F_1$, $F_2$, and $F_3$ are shown in Fig.~\ref{fig:uncertain_queueing_problem:indended} for selected egress ports where at least two streams are contesting for resources:
The downward-facing arrows denote the time of enqueuing a frame of the shown stream in the respective egress queue and the boxes depict the transmission slots where the gate is opened and a frame can be transmitted (as determined by the GCL).

The problem of employing conventional TSN schedulers directly is that they assume near-deterministic delay characteristics, e.g., the arrival time of $F_1$ and $F_2$ at the NW-TT is strictly periodic with minor delay variations in the range of microseconds.
In contrast, the 6G delay variations are much larger and can cause $F_2$ to arrive earlier than $F_1$ at the NW-TT.
The resulting QoS impairments are illustrated in Fig.~\ref{fig:uncertain_queueing_problem:faulty}:
First, $F_2$ steals the intended transmission slot of $F_1$, causing $F_1$ to miss its slot at the subsequent hop $[B_1, B_2]$ and having to wait for the next transmission opportunity in the next hypercycle.
This may already be enough to violate the end-to-end latency requirement of $F_2$.
Second, $F_2$ arrives at $B_1$ earlier than expected and also steals the intended transmission slot of $F_3$.
This demonstrates that the queuing reordering effects can cascade to the wired TSN backbone and potentially result in the same QoS violations for wired streams as for $F_1$.

\textbf{Problem Statement:}
In this paper, we aim to address the root of this issue by compensating the significant 6G packet delay variations in 6G-TSN bridges.
We do so by extending the hold-and-forward buffering mechanism of 5G~\cite{3gpp.23.501} with a de-jittering functionality which can be realized with any of the different approaches that we call \textit{Packet Delay Correction~(PDC)}.
Compared to previous approaches that aim to enable TSN scheduling in 5G-TSN settings~\cite{9212049,9940254,Egger2025}, we stress that PDC does not require special scheduling constraints but rather acts as a complementary solution to enable the plethora of conventional TSN schedulers from wired TSN.

%% file: content/04_design.tex
\section{Packet Delay Correction}

PDC is a concept that aims at modifying the per-frame delay in order to obtain a low  PDV. To this purpose, PDC  relies on the hold-and-forward buffering mechanism (HF) at the traffic egress point implementing a new function that corrects the deviation between a target delay ($target\_delay$) and the delay experienced by a frame within the 6G system $d$ (a.k.a. residence time). The main idea is to hold every frame until it reaches a predefined $target\_delay$. By holding back frames, the average delay increases. Hence, PDC provides a trade-off between longer (but deterministic) frame delays with low  PDV and a low average delay with a large PDV. For most time-critical applications, high average delay is not as important as a low PDV, since it is sufficient that the frames arrive before their maximum delay bound.

\subsection{PDC based on Timestamps}
\begin{figure}
    \centering
    \includegraphics[width=\linewidth]{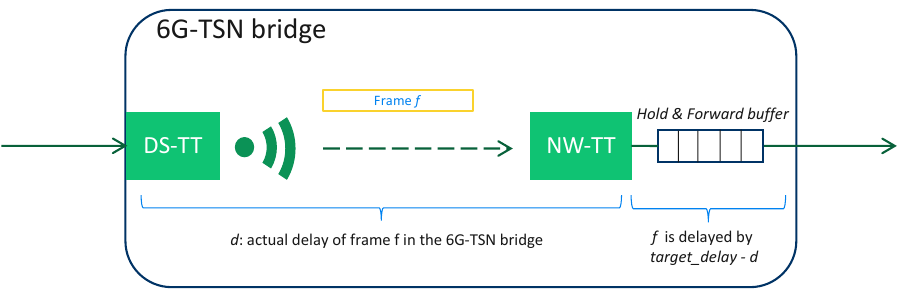}
    \caption{Packet delay correction mechanism}
    \label{fig:pdc}
\end{figure}

One solution is to timestamp a frame f at ingress and egress of the 5G system, i.e., at DS-TT (ingress) and NW-TT (egress) for uplink as shown in Fig. \ref{fig:pdc}, at NW-TT (ingress) and DS-TT (egress) for downlink, or at DS-TT (ingress) and another DS-TT’(egress) for UE to UE communications.
When the delay experienced by a frame within the 6G system $d$ is known (by subtracting egress and ingress timestamps), then  $target\_delay - d$  is the time to hold the frame in the HF buffer to reach $target\_delay$ (see Fig. \ref{fig:pdc}). The timestamp included in the frame shall be removed from the frame before it is forwarded to the next node after the 6G-TSN bridge. This approach requires that the DS-TTs and NW-TT are synchronized with the same (6G) clock \cite{9204594}. 

It has to be noted that wire-speed timestamping can be a  challenging task. Some programmable packet processors \cite{highperformance_packet_timestamping} could be used in the DS-TT and NW-TT for this task, but this would come at an additional cost.

Another consideration is how the timestamp is transported in every frame. A timestamp requires at least 8 bytes in the header of the frame. It is not desirable to standardize changes in the Ethernet headers for this purpose. One solution could be to add such header field to include the timestamp to the 3GPP transport protocols. However, this would require 3GPP standardization. 

\subsection{PDC based on Virtual Time Slots}
We propose a solution that uses virtual time slots instead of precise (hardware) timestamps; this approach also requires that both NW-TT and DS-TT are synchronized with a common (6G) clock. Virtual time slots are defined as a very short time intervals (with a size in the order of microseconds) that is identified with an integer which we call the slot ID.
A virtual time slot is basically a synchronized counter in the 6G system that is used in the NW-TT and DS-TT. It counts time steps of a certain length that provide sufficient precision for measuring the packet delay, e.g. in the order of several microseconds, but with a much coarser precision than a hardware timestamp. In other words, virtual time slots provide a relaxed variant of timestamping: they have lower time precision than timestamps (but sufficient to measure packet delays) and are smaller in size than a timestamp, they do not represent any global time but are based on 6G system internal variables for measuring delays, and they can be efficiently implemented. 

\begin{figure}
    \centering
    \includegraphics[width=\linewidth]{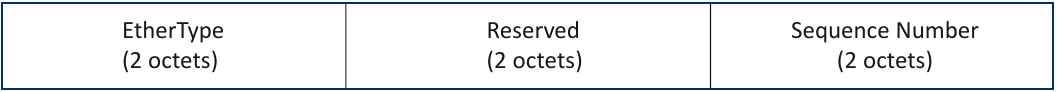}
    \caption{R-TAG format, as defined in IEEE 802.1CB \cite{802.1CB}}
    \label{fig:rtag}
\end{figure}

When TSN frames arrive to the 6G network, they need to be marked with their virtual time slot. We propose to carry the virtual time slot in the Ethernet header. For this, we suggest re-using the Ethernet R-Tag (as defined in IEEE 802.1CB ~\cite{802.1CB}) in  the Ethernet frame header for transferring frame timing information.
The “Sequence Number” field, as shown in Fig. \ref{fig:rtag}, can be used to carry the value of the virtual time slot: the slot ID. Slot ID refers to the arrival slot that is encoded in the R-Tag. 
Nothing prohibits the use of multiple R-Tags in an Ethernet frame.
In this case, an R-Tag already defined for other purposes can be reused to insert a virtual time slot. This additional R-Tag is used only within the 6G system.
It is added at ingress and removed at egress TT.
Therefore, the R-Tag used inside the 6G system is not visible to the outside world. Ethernet frames may use other R-Tag(s) for which the 6G system is transparent, and hence these R-Tag(s) remain untouched by the 6G system.

The tasks at the ingress TT are: (i) identify the time slot with specific slot ID when the frame arrives, (ii) encode slot ID in the added R-Tag (as Sequence Number).
At the egress TT the tasks are: (i) identify the egress slot ID, (ii) based on R-Tag and predefined $target\_delay$ within the 6G system for this stream, calculate the time slot at which the frame should be transmitted at egress ($slotID + target\_delay$), (iii) buffer the frame until the calculated time slot in (ii), and (iv) remove R-Tag and transmit. In this way, all frames (belonging to the same traffic class) will experience the same target delay within the 6G system, and therefore resulting in a small delay variation.

\begin{figure}
    \centering
    \includegraphics[width=\linewidth]{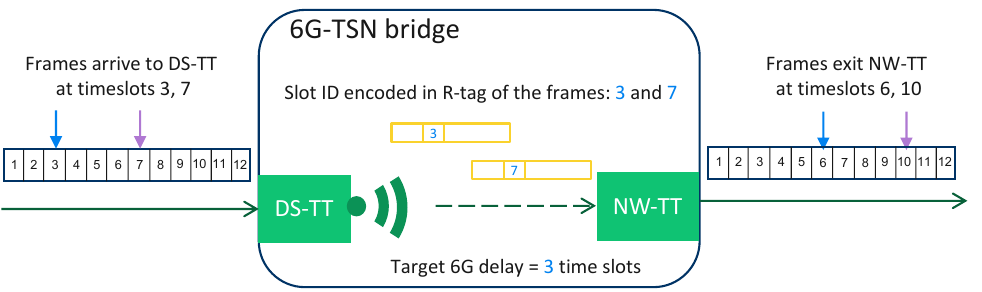}
    \caption{Example of PDC based on virtual time slots.}
    \label{fig:pdc_timeslots}
\end{figure}

In Fig. \ref{fig:pdc_timeslots}, the use of the virtual time slots is shown using an example.
Assuming that there are two frames arriving at timeslots with slot ID 3 and 7,respectively, the ingress TT (DS-TT) will encode the slot ID in each respective R-tag field of the frames.
When the frames arrive at the egress TT (NW-TT), the $target\_delay$ (3 slots) is added to the encoded slot ID of every corresponding frame in order to obtain the egress time slots: 6 and 10, respectively.

\subsection{Configuration Options for PDC} \label{sec:pdc_config}
\begin{figure}[t]
    \centering
    \begin{subfigure}{0.48\textwidth}
        \centering
	\includegraphics[width=\textwidth]{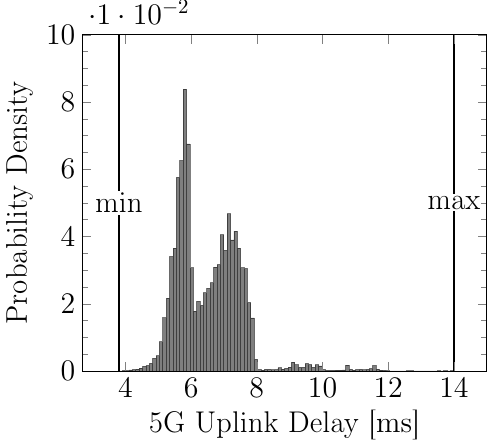}
    \end{subfigure}
    \hfill
    \begin{subfigure}{0.48\textwidth}
        \centering
	\includegraphics[width=\textwidth]{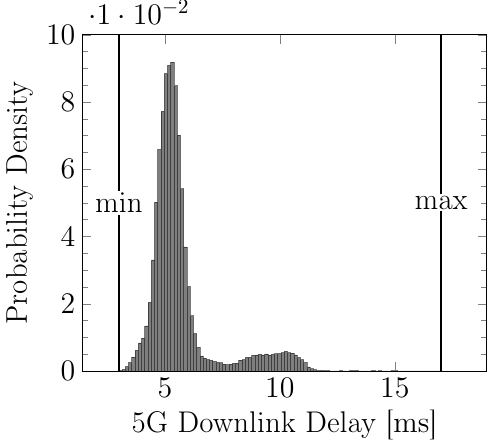}
    \end{subfigure}
    \caption{5G uplink/downlink delay histograms, measured by \cite{downlink_example_histogram}.}
    \label{fig:5gdelay}
\end{figure}

The value of $target\_delay$ can be obtained based on an estimation of the range of latencies that are observed in the 5G/6G network, e.g., via data-driven analysis such as latency estimation as described in \cite{9708928}.
$target\_delay$  can also be estimated via an expert system that uses historical data on other deployments.
Another possibility is to set $target\_delay$ value to the packet delay budget (PDB) value configured for the flow in the 6G system.
In this paper, we use real histograms captured from measurements of a 5G network as shown in Fig. \ref{fig:5gdelay}, where $max$ equals an upper bound of the delay for example provided by URLLC.

When $target\_delay$ is configured to equal $max$, the delay of all frames is compensated such that their resulting delay equals $max$.
In this case, the achieved reliability is equivalent to the statistical confidence that the delay of frames never exceed $max$.
However, it is also possible to configure $target\_delay$ to be smaller than the expected maximum possible delay $max$.
In this case, frames with a greater delay than $target\_delay$ can be filtered out at the egress TT, as specified in \cite{3gpp.23.501}, clause 5.7.3.4.
This allows to provide smaller end-to-end latency guarantees to applications with lower reliability requirements.
For example, the stream of an application only requiring a reliability of \qty{90}{\percent} can be configured to have a $target\_delay = 0.9 \cdot max$ cutting of the long tail of the distributions.

\subsection{Simulator Implementation of PDC}
\label{sec:design:impl}
To be able to analyze the effect of PDC in converged 6G-TSN networks, we provide a simulator-based implementation of PDC.
Our implementation is based on the 6GDetCom Simulator \cite{Haug2025} for converged 6G-TSN networks, which builds on the OMNeT++ Simulator \cite{Varga2010} with the open-source INET Framework \cite{Mros2019}.
The 6GDetCom Simulator provides a 6G-TSN bridge including an NW-TT and multiple DS-TTs according to the 3GPP support for TSN specified in \cite{3gpp.23.501}.
This 6G-TSN bridge is parametrized to represent 6G network characteristics while using 5G network performance measurements.
For example, it can be configured to delay frames according to delay distributions from measurements in real-world testbeds as shown in Fig. \ref{fig:5gdelay}.
We provide a detailed description of our PDC implementation in the following.

In order to simulate PDC based on timestamps, we implement a \textit{timeTagging} module at the DS-TTs and NW-TT of the 6G-TSN bridge in the 6GDetCom Simulator.
This \textit{timeTagging} module adds a tag with the ingress timestamp $T_i$ to every incoming frame at the ingress TT.
At the following (egress) TT, the \textit{timeTagging} module generates an egress timestamp $T_e$ and together with the received ingress timestamp $T_i$ calculates the delay experienced by the frame inside the 6G-TSN bridge $d=T_e - T_i$ (a.k.a. residence time).

Furthermore, we provide a \textit{pdc} module that allows to specify a $target\_delay$ value per stream, as well as a default value for all other streams.
The identification of streams is performed by the already existing \textit{streamIdentifier} of INET, which allows to identify streams based on various parameters, such as their source and destination nodes, Priority Code Point (PCP) values and more.
An example configuration is shown in Listing \ref{code:pdc_timestamp}.
The \textit{pdc} module then uses this value to hold back the frame for $target\_delay - d$ before releasing it on the egress port, as described above.

\begin{listing}[h]
	\begin{minted}{ini}
# per stream pdc
*.detCom.nwtt.bridging.pdc.mapping = [{stream: "stream1", pdc: "10ms"}]

# default pdc
*.detCom.nwtt.bridging.pdc.defaultPdc = 15ms
	\end{minted}
	\caption{Example time-stamp based PDC configuration}
	\label{code:pdc_timestamp}
\end{listing}

PDC for the uplink direction is configured on the NW-TT, while PDC for the downlink direction is performed on the DS-TT of the corresponding device.

For PDC based on virtual time slots, a frame is not forwarded after a specific time, but within a specified timeslot.
This effectively means an additional delay variation equal to the size of the time slots is introduced.
For example, if a frame is scheduled for a timeslot $x$ of size $\qty{100}{\us}$, the frame can be transmitted anywhere within this interval of $\qty{100}{\us}$.
To this end, our PDC implementation allows specifying an additional \textit{jitter}-parameter to simulate the behavior of virtual time slots.
An example is provided in Listing \ref{code:pdc_timestamp}. 

\begin{listing}[h]
	\begin{minted}{ini}
# per stream pdc for a timeslot of size 100 us
*.detCom.nwtt.bridging.pdc.mapping = [{stream: "stream1", pdc: "10ms",  jitter: "uniform(0ms,100us)"}]
	\end{minted}
	\caption{Example virtual time slot based PDC configuration}
	\label{code:pdc_timeslot}
\end{listing}

%% file: content/05_evaluation.tex
\section{Evaluation}

In the following, we evaluate PDC by means of the 6GDetCom Simulator as described above.
We demonstrate that PDC enables conventional  TSN scheduling techniques to be applied in 6G-TSN networks. 
Moreover, we evaluate the impact of different slot sizes for the  virtual time slots. 

\begin{figure}
    \centering
    \includegraphics[width=0.8\textwidth]{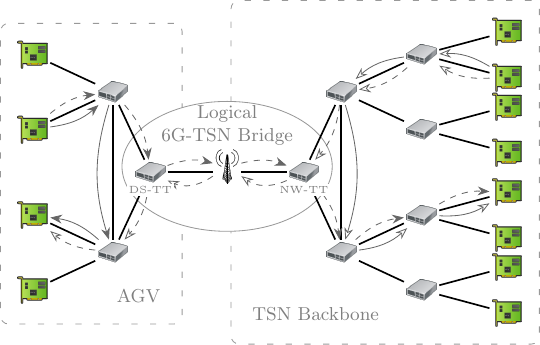}
    \caption{Simulation Setup}
    \label{fig:simsetup}
\end{figure}

\subsection{Methodology} \label{sec:eval:methodology}
For our evaluation, we consider the scenario shown in Fig. \ref{fig:simsetup}, where an automated guided vehicle (AGV) is connected to the backbone TSN network of a factory.
The logical 6G-TSN bridge is connecting two wired TSN network segments: The AGV-internal network on the device side is connected to the DS-TT and the TSN backbone is connected to the NW-TT.

For the evaluation, we differentiate between wired streams and wireless streams.
Wired streams have their source and destination in the same part of the network (i.e. they do not pass through the 6G-TSN bridge).
Wireless streams have their source and destination in different parts of the network (i.e. they pass through the 6G-TSN bridge).
The parameters of both streams are given in Table \ref{tab:streams}.
Both types of streams are configured to have the same Priority Code Point (PCP) value of $7$.
All Ethernet links in the network are configured to have a constant link speed of $\qty{100}{Mbps}$ and a propagation delay of $\qty{50}{\ns}$.
The wireless link between the DS-TT and NW-TT is configured using the uplink and downlink histograms as shown in Fig. \ref{fig:5gdelay}.

\begin{table}[]
    \centering
    \begin{tabular}{c|c|c|c|c|c}
         & \textbf{Period} & \textbf{Frame size} & \textbf{Latency} & \textbf{Jitter} & \textbf{PCP} \\\hline
         \textbf{Wired Streams} & $\qty{5}{\ms}$ & $\qty{100}{B}$ & $\qty{5}{\ms}$ & $\qty{1}{\us}$ & $7$ \\
         \textbf{Wireless Streams} & $\qty{20}{\ms}$ & $\qty{100}{B}$ & $\qty{20}{\ms}$ & $\qty{100}{\us}$ & $7$ \\
    \end{tabular}
    \caption{Stream Parameters}
    \label{tab:streams}
\end{table}

Our scheduler implementation is based on the TSN scheduling constraints of~\cite{Craciunas2016RTNS}.
As PDC is not specific to any specific TSN scheduler, however, we argue that the results stay quantitatively the same for other scheduling constraints (e.g., ~\cite{8894249, nwps}). 
The output of this scheduler is used to configure the GCLs in a simulated network based on Fig. \ref{fig:simsetup}.
We perform our evaluation based on the simulation PDC implementation as described in \ref{sec:design:impl}.

\subsection{Effect of PDC on TSN Traffic scheduling}

We start by demonstrating that PDC enables the usage of conventional TSN scheduling techniques in 6G-TSN settings.
To this end, PDC is compared with joint 6G-TSN configurations that rely on known constant wireless delay characteristics (e.g., as in ~\cite{9212049,9940254}).
In particular, we compare the simulated end-to-end latency bounds of a conventional TSN scheduler under two different 6G channel assumptions that rely on
\begin{itemize}
    \item \textit{median} 6G delays of $\qty{6.38}{\ms}$ and $\qty{5.26}{\ms}$ for uplink and downlink streams,
    \item \textit{maximum} 6G delays of $\qty{14}{\ms}$ and $\qty{17.1}{\ms}$ for uplink and downlink streams
\end{itemize}
based on the latency measurements from a 5G testbed as presented in Fig. \ref{fig:5gdelay}.
The following uses \textsc{Med} and \textsc{Max} to denote the above 6G channel assumptions, respectively.
The two simulation scenarios under these channel assumptions use a statically configured GCL provided by the conventional schedular without the usage of PDC.
In contrast, \textsc{PDC} configures $target\_delay$ in accordance with the maximum 6G delays in listed in the \textsc{Max} channel assumption. This compensates the 6G delay variations and provides latencies around the $target\_delay$ value according to the virtual time slot size of PDC. 
The IEEE 802.1Qbv GCLs are computed for a stream set with 50 uplink and 50 downlink streams, along with 5 internal streams per wired partition.
Their stream parameters are defined in Section~\ref{sec:eval:methodology}.
The simulation is repeated for more than 1 million transmission cycles per stream to retrieve meaningful insights for streams with ultra-high reliability requirements of up to $\qty{99.999}{\percent}$.

\begin{figure}
    \centering
    \begin{subfigure}{0.48\textwidth}
        \centering
        \includegraphics[width=\textwidth]{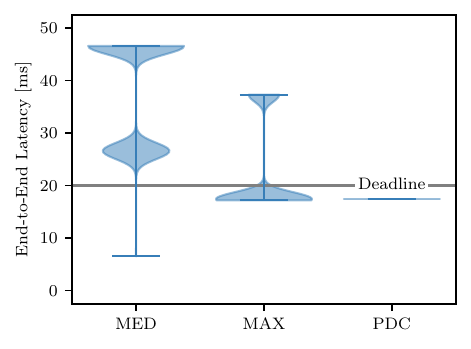}
        \caption{Wireless streams.}
        \label{fig:res_dist:wireless}
    \end{subfigure}
    \hfill 
    \begin{subfigure}{0.48\textwidth}
        \centering
        \includegraphics[width=\textwidth]{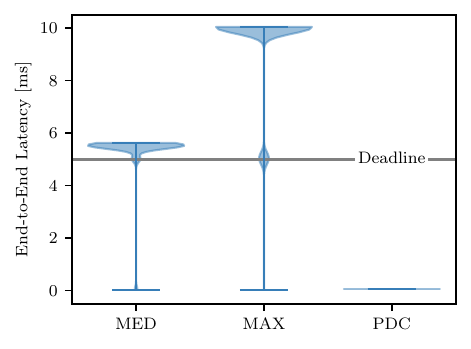}
        \caption{Wired streams.}
        \label{fig:res_dist:wired}
    \end{subfigure}
    \caption{End-to-end latency probability distribution in the form of violin plots for \textsc{Med}, \textsc{Max}, and \textsc{Pdc} for wireless and wired streams respectively.
    }
    \label{fig:res_dist}
\end{figure}

Fig.~\ref{fig:res_dist} shows the simulation results for the wired and wireless stream with the worst end-to-end latency that was observed.
First and foremost, the results clearly show that, for \textsc{Med} and \textsc{Max}, there is a fundamental divergence between the 6G delay assumptions for synthesizing the IEEE 802.1Qbv schedule and the actual 6G packet delay variations that are observed at runtime.
This divergence leads to frames taking transmission slots intended for the transmission of different frames which can effectively nullify any supposed end-to-end latency guarantees.
Moreover, Fig.~\ref{fig:res_dist} also shows that this not only affects wireless streams, but potentially also wired streams.
Section~\ref{sec:eval:in-depth-analysis} analyzes the underlying cause for these cascading effects further.
In contrast, the results show that \textsc{Pdc} eliminates this divergence and achieves bounded end-to-end latency and jitter guarantees for both wired and wireless streams.
Thereby, PDC enables the usage of conventional scheduling techniques that are well-established from wired TSN.

\begin{figure}[t]
    \centering
    \begin{subfigure}{0.93\textwidth}
        \centering
        \includegraphics[width=\textwidth]{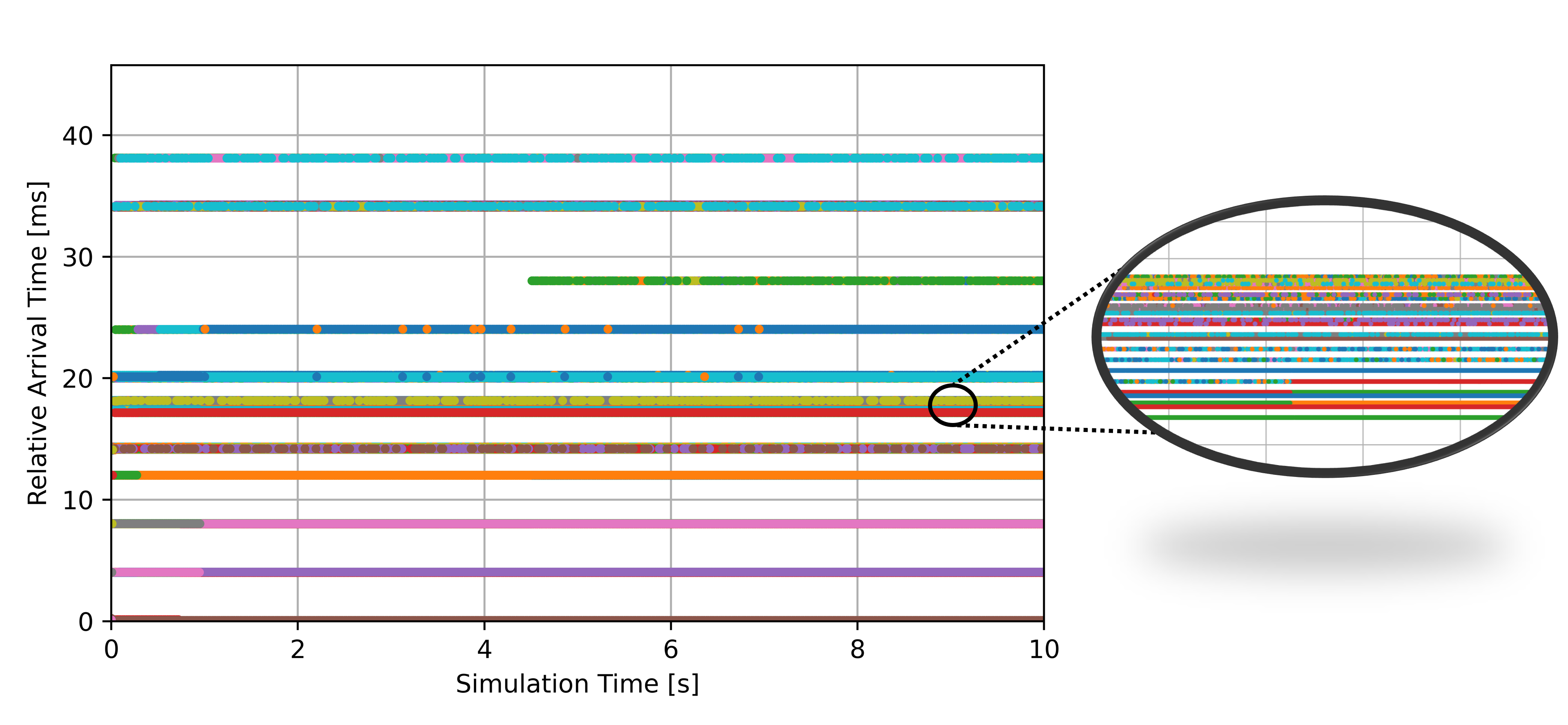}  
        \caption{Without PDC}
        \label{fig:in_depth:nopdc}
    \end{subfigure}
    \begin{subfigure}{0.93\textwidth}
        \centering
        \includegraphics[width=\textwidth]{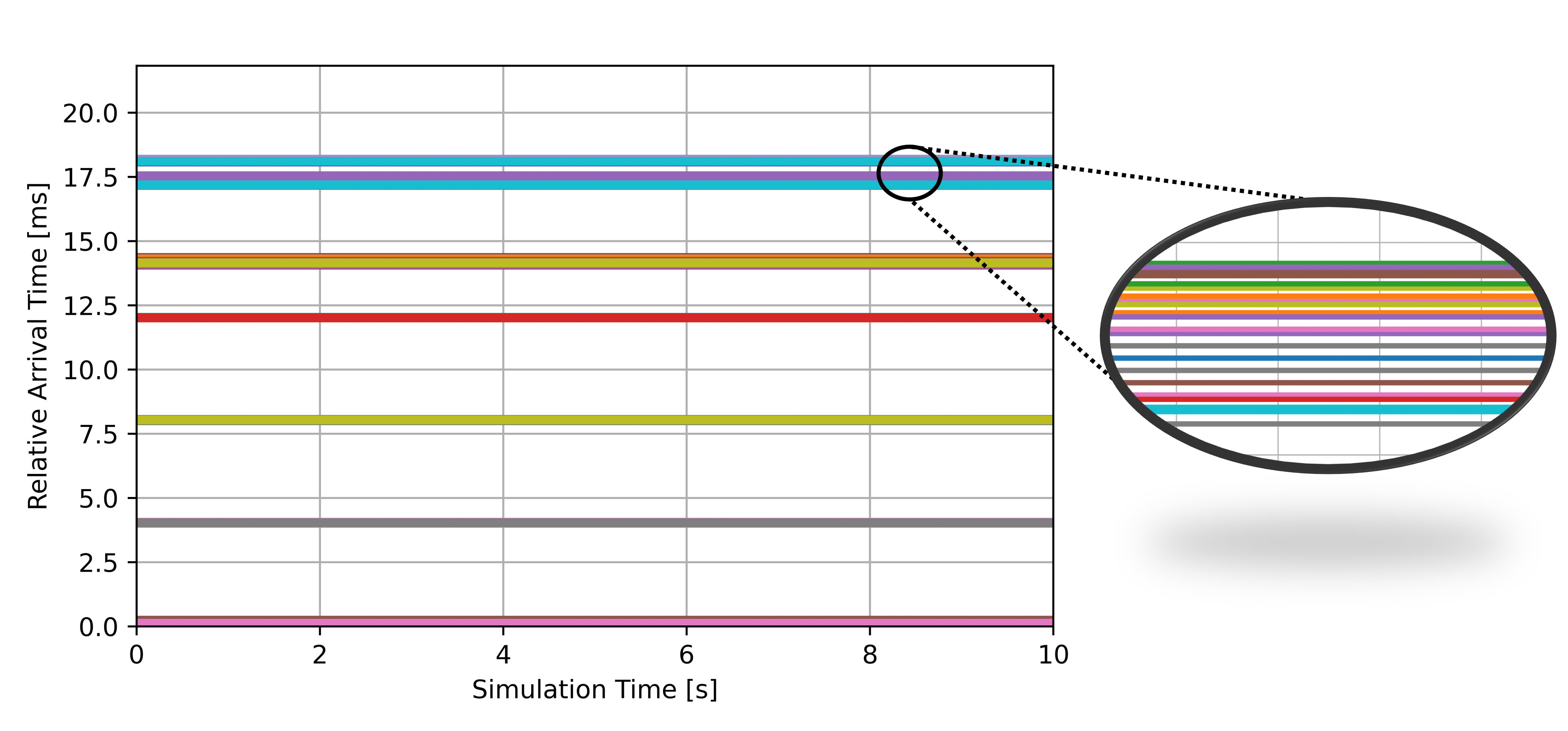}  
        \caption{With PDC}
        \label{fig:in_depth:pdc}
    \end{subfigure}
    \caption{Transmission behavior of scheduled traffic under the Max 6G delay assumption with and without PDC. Different colors correspond to different TSN streams.}
    \label{fig:in_depth}
\end{figure}

\subsection{In-Depth Analysis: The Uncertain Enqueueing Problem} \label{sec:eval:in-depth-analysis}

While Fig.~\ref{fig:res_dist} shows the experienced latency for a single stream, the underlying causes for cascading deadline violations are best seen from a global view of the complete set of streams.
Fig. \ref{fig:in_depth} provides this view by showing the relative arrival time of each stream (wired and wireless) at its listener within a simulated time window of $\qty{10}{\s}$.
Each color corresponds to the arrival times of a single stream.
IEEE 802.1Qbv intends to schedule traffic so that the traffic of each stream arrives at the receiver at a determined arrival time which would correspond to a straight horizontal line for every stream in Fig. \ref{fig:in_depth}.
Any deviation from the intended IEEE 802.1Qbv schedule is thereby shown by fragmented arrival times instead of constant arrival times.

For the case of \textsc{Max} 6G delay assumptions without PDC, this effect is clearly captured by Fig.~\ref{fig:in_depth}a:
The 6G packet delay variations can cause two frames \ldash say $f_1$ and $f_2$ of two different uplink streams \rdash to be enqueued in a non-deterministic order at the NW-TT following the 6G transmission.
That is, although the scheduler might assume that $f_1$ always arrives at the NW-TT before $f_2$, this assumption can be violated if the actual 6G delay experienced by $f_2$ is significantly less than what is assumed by \textsc{Max}.
Any reordering effect of this sort entails that $f_1$ would experience additional queuing delays that can cause $f_1$ to miss its intended GCL transmission slot.
Fig.~\ref{fig:in_depth:nopdc} demonstrates that this is not a rare event as can be seen by the scattered arrival times of different frames of a single TSN stream.
Rather, it can be seen that it is a quite common situation.
One consequence of this is that frames miss their intended TSN schedules and, as a result, the transmission is deferred by one additional hypercycle and thereby exceed the maximum delay bound of \qty{20}{\ms}.
In contrast, when PDC is applied it removes delay uncertainties and the TSN scheduler manages to scheduling TSN streams to a target slot and the maximum delay bound is maintained, as seen in Fig. \ref{fig:in_depth:pdc}.

\subsection{Impact of Virtual Time Slots}
Next, we evaluate the impact of different virtual time slots to realization for PDC.  
To quantify the impact of different granularity levels of virtual time slots, we analyze their scalability in the number of schedulable streams. 
That is, using the same network topology and stream specifications as before, we evaluate the maximum number of wireless streams (out of a randomly generated stream set of 200 uplink and downlink streams) for which a feasible 802.1Qbv schedule is found. 
We consider a fixed number of 15 wired streams per wired network partition for this evaluation.
The experiments are performed for virtual slot sizes between $\qty{1}{\us}$ and $\qty{500}{\us}$.

\begin{figure}
    \centering
    \includegraphics[width=\textwidth]{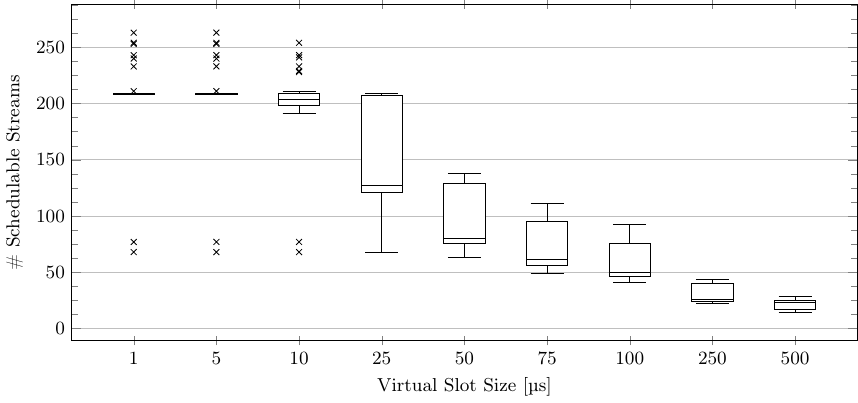}
    \caption{Scalability impact of virtual time slot sizes.}
    \label{fig:res_timeslots}
\end{figure}

Fig. \ref{fig:res_timeslots} shows the results in the form of boxplots for 1000 randomly generated stream sets showing the minimum, median, and maximum number of scheduled streams per streamset, as well as, lower and upper quartiles.
It shows a clear downward trend in the number of schedulable streams when the virtual slot size increases.
This is expected, as larger virtual time slots essentially reduce the precision of PDC and thereby create an own packet delay variation, which is bounded to the virtual time slot length.
To illustrate, consider the maximum 6G uplink and downlink delay of $\qty{14}{\ms}$ and $\qty{17.1}{\ms}$ from the utilized histograms~\cite{downlink_example_histogram} and a virtual time slot size of $\qty{500}{\us}$.
As PDC remains a delay uncertainty of $\qty{500}{\us}$, the TSN scheduler can only clearly separate 12 uplink and 5 downlink streams per TSN egress queue, while staying below a maximum delay of $\qty{20}{\ms}$.

What Fig. \ref{fig:res_timeslots} also shows is that virtual time slots with a granularity below $\qty{10}{\us}$ are not limiting the TSN capacity in the network.
But as the virtual time slot granularity grows larger, the resulting delay uncertainty limits the TSN scheduler’s ability to isolate multiple streams, thereby reducing the number of schedulable streams.

%% file: content/06_discussion.tex
\section{Discussion}

Before concluding this paper, this section revisits the underlying design choices of PDC to discuss and reflect on its theoretical and practical implications.

\textbf{Configuring Target 6G Delays.} 
There is an apparent trade-off that needs to be considered when configuring the \textit{target\_delay} parameter for PDC.
Choosing the target delay too large results in a higher queuing delays and higher buffer requirements at the egress side of the 6G systems.
Conversely, choosing the target delay too small increases the probability that the actual 6G packet delays exceed the configured \textit{target\_delay} and can result in similar violations of the IEEE 802.1Qbv schedule as described in Section~\ref{sec:uncertain_queueing}.
We argue that 6G delay histograms can enable a suitable trade-off, as detailed in Section~\ref{sec:pdc_config}: 
On a per-stream basis, the stream's reliability requirement can be used to compute the corresponding 6G histogram percentile and to jointly configure PDC and the TSN schedule.
Still, any packet experiencing a 6G delay that exceeds this value must be strictly discarded/degraded to avoid interference with other TSN streams.

\textbf{Comparison to Other Traffic Shapers.}
In this work, we focused on the impact of large 5G/6G packet delay variations on scheduled traffic with IEEE 802.1Qbv.
We argue, however, that the impact on other traffic shapers in TSN \ldash e.g., strict priority, the credit-based shaper (CBS), or the asynchronous traffic shaper (ATS) \rdash can have similar, albeit more subtle effects.
In particular, the above shapers rely on analytical techniques to compute per-hop queuing delay bounds and to determine certain configuration parameters (e.g., the maximum residence time in case of the ATS).
Without accounting for the probabilistic 6G delay characteristics, the end-to-end guarantees of these analytical techniques are nullified in the same manner as we showed for the validity of IEEE 802.1Qbv schedules.
Due to the complementary design of PDC, we argue that it can pose a simple mitigation strategy that allows employing wireless-unaware traffic engineering in TSN.

%% file: content/07_conclusion.tex
\section{Conclusion}
With 5G, support has been introduced into mobile networks to provide upper delay bounds when providing connectivity for time-critical applications. Nevertheless, communication in 5G mobile networks - and more generally in wireless networks - introduce significant packet delay variation (PDV). For wired local area networks, time-sensitive networking (TSN) has been standardized to provide lossless, dependable communication for delay-bounded and time-critical applications. To this end, a TSN controller performs (IEEE 802.1Qbv) time scheduling for TSN traffic throughout the network; this allows to reserve transmission opportunities for time-critical TSN streams in the network, in order to guarantee timely delivery of data. The benefits and flexibility of wireless connectivity have sparked the interest to extend TSN communication over wireless networks. To this end, support for TSN has been standardized for 5G, so that the 5G system appears as a logical TSN node within the TSN network.

It has been shown in \cite{Egger2025} that PDV of TSN nodes poses a challenge to TSN scheduling; the uncertainty of frame arrivals due to PDV makes it difficult to schedule subsequent hops.
We propose a method of packet delay correction (PDC), which introduces into a 5G or future 6G network the capability to remove PDV by means of a hold-and-forward buffering mechanism at the egress point of the mobile network (i.e. at the device side or network side TSN translators).
The de-jittering functionality of PDC is achieved by determining the residence time for every frame through the 5G/6G network and releasing them from an egress queue after a configured target delay has been reached. We propose that virtual time slots are used for determining the residence time and releasing frames at the egress. Virtual time slots can be implemented efficiently without a costly need for precise timestamping at line rate. Virtual time slots are configurable time steps according to which PDC is performed. The virtual slot size defines the granularity of PDC; it also defines the remaining PDV of the 5G/6G network. 

We evaluate PDC via simulations of TSN scheduling in a TSN network that integrates wired TSN with 5G/6G.
Firstly, our evaluation demonstrates that PDC facilitates the integration of conventional TSN scheduling techniques within converged 6G-TSN networks.
This is a relevant result that emphasizes PDC as an enabler for seamless TSN scheduling that does not need to be adapted to the complexity of the relatively high PDV in the 6G-TSN bridges.
Regarding the PDC based on virtual time slots, we show that for large virtual slot sizes in the 100's of microseconds, the uncertainty of frame delivery complicates the TSN scheduling and results in low TSN network capacity in terms of the number of schedulable TSN streams.
As the virtual slot size decreases, PDC in a 5G/6G system can increase the capacity of the scheduled TSN network significantly.
For virtual slot sizes below 10 microseconds only marginal further gains are achievable.
The results indicate that low-jitter connectivity, based on virtual time slot based PDC, is a promising capability for future 6G networks to provide high performance and capacity in future TSN networks.